# Single-shot 3D structure determination of nanocrystals with femtosecond X-ray free electron laser pulses

Rui Xu[1], Huaidong Jiang[2], Changyong Song[3], Jose A. Rodriguez[4,5], Zhifeng Huang[1], Chien-Chun Chen[1], Daewoong Nam[3,7], Jaehyun Park[3], Marcus Gallagher-Jones[3,8], Sangsoo Kim[3], Sunam Kim[3], Akihiro Suzuki[9], Yuki Takayama[3,6], Tomotaka Oroguchi[3,6], Yukio Takahashi[9], Jiadong Fan[2], Yunfei Zou[1], Takaki Hatsui[3], Yuichi Inubushi[3], Takashi Kameshima[10], Koji Yonekura[3], Kensuke Tono[10], Tadashi Togashi[10], Takahiro Sato[3], Masaki Yamamoto[3], Masayoshi Nakasako[3,6], Makina Yabashi[3], Tetsuya Ishikawa[3] and Jianwei Miao[1*]

*[1]Department of Physics and Astronomy, and California NanoSystems Institute, University of California, Los Angeles, CA 90095, USA. [2]State Key Laboratory of Crystal Materials, Shandong University, Jinan 250100, China. [3]RIKEN SPring-8 Center, Kouto 1-1-1, Sayo, Hyogo 679-5148, Japan, 1-1-1, Kouto, Sayo, Hyogo 679-5148, Japan. [4]Department of Biological Chemistry, UCLA-DOE Institute for Genomics and Proteomics, University of California, Los Angeles, California, 90095, USA. [5]Howard Hughes Medical Institute (HHMI), Chevy Chase, Maryland 20815-6789, USA. [6]Department of Physics, Faculty of Science and Technology, Keio University, 3-14-1 Hiyoshi, Kohoku-ku, Yokohama, Kanagawa 223-8522, Japan. [7]Department of Physics, Pohang University of Science and Technology, Pohang 790-784, Korea. [8]Institute of Integrative Biology, University of Liverpool, Liverpool L69 7ZB, United Kingdom. [9]Department of Precision Science and Technology, Graduate School of Engineering, Osaka University, 2-1 Yamada-oka, Suita, Osaka 565-0871, Japan. [10]Japan Synchrotron Radiation Research Institute, Kouto 1-1-1, Sayo, Hyogo 679-5198, Japan.* *e-mail: miao@physics.ucla.edu

**Coherent diffraction imaging (CDI) using synchrotron radiation, X-ray free electron lasers (X-FELs), high harmonic generation, soft X-ray lasers, and optical**

lasers has found broad applications across several disciplines[1-20]. An active research direction in CDI is to determine the structure of single particles with intense, femtosecond X-FEL pulses based on diffraction-before-destruction scheme[13-15,21-25]. However, single-shot 3D structure determination has not been experimentally realized yet. Here we report the first experimental demonstration of single-shot 3D structure determination of individual nanocrystals using ~10 femtosecond X-FEL pulses. Coherent diffraction patterns are collected from high-index-faceted nanocrystals, each struck by a single X-FEL pulse. Taking advantage of the symmetry of the nanocrystal, we reconstruct the 3D structure of each nanocrystal from a single-shot diffraction pattern at ~5.5 nm resolution. As symmetry exists in many nanocrystals and virus particles[26,9,15,27], this method can be applied to 3D structure studies of such particles at nanometer resolution on femtosecond time scales.

The X-FEL experiment was conducted using the SPring-8 Angstrom Compact Free-Electron Laser (SACLA)[28]. Figure 1 shows a schematic layout of the experimental setup. X-FEL pulses (photon energy: 5.4 keV, pulse duration: ~10fs[29], and pulse energy: ~100 μJ) were focused to a ~1.5 μm spot by a pair of Kirkpatrick-Baez (K-B) mirrors[30]. The focused X-FEL pulses impinged on high-index-faceted gold nanocrystals deposited on a 100-nm-thick $Si_3N_4$ membrane. The gold nanocrystals, synthesized by the seed-mediated growth method to ensure monodisperse trisoctahedral shape and size (Supplementary Methods and Fig. S1), consist of a high density of low-coordinated atoms such as steps, edges and kinks, serving as catalytically active sites[26]. By scanning the $Si_3N_4$ membrane relative to the focused X-FEL pulses, coherent X-ray diffraction patterns were collected on a multi-port charge coupled device detector, each generated by a single pulse based on the diffraction-before-destruction scheme[21,13,15] (Methods).
Supplementary video S1 shows a typical single-shot X-FEL experiment. The density of
nanocrystals deposited on the $Si_3N_4$ membrane was tuned to control the hit-rate for
these experiments. Supplementary Fig. 2 shows high-index-faceted gold nanocrystals
deposited on a $Si_3N_4$ membrane before and after being struck by X-FEL pulses. Using
this data acquisition scheme, we collected terabytes of coherent X-ray diffraction
patterns. After screening and post analysis of this large amount of data (Methods), we
selected 100 good quality diffraction patterns from single isolated and highly
symmetrical gold nanocrystals (Supplementary video S2). The best six single-shot
diffraction patterns are shown in Fig. 2a and Supplementary Fig. S3, whose diffraction
signal is presently limited by the detector size.

　　　Next, we verified the trisoctahedral symmetry of the high-index-faceted gold
nanocrystals based on the single-shot diffraction patterns. Trisoctahedral nanocrystals
belong to the crystallographic point group $m\bar{3}m$, consisting of 2-, 3- and 4-fold
rotational symmetry along the <011>, <111> and <001> directions, respectively[31] (Fig.
2b). Given a single-shot diffraction pattern of a trisoctahedral nanocrystal sampled on
the Ewald sphere, there are 48 symmetry operations that can generate 48 spherical
diffraction patterns[31]. These diffraction patterns intersect with each other to form 14
independent pairs of self-common arcs (Supplementary Methods). By analyzing the six
single-shot diffraction patterns, we confirmed the existence of 14 independent pairs of
self-common arcs in each diffraction pattern. Figure 2c and Supplementary Fig. S4
show four representative pairs of the self-common arcs for Fig. 2a. We quantified the
self-common arcs for the six single-shot diffraction patterns by using $R_{arc}$
(Supplementary Methods). The average $R_{arc}$ for the six diffraction patterns is 5.1%,

6.5%, 8.1%, 8.3%, 8.7%, and 9.1%. These values are smaller than those used in single particle cryo-electron microscopy (EM) of viruses[27], suggesting that these single-shot diffraction patterns have high signal to noise ratios, and that our screening and post data analysis approach allows us to select single-shot diffraction patterns from highly symmetrical nanocrystals.

After verifying the trisoctahedral symmetry of the gold nanocrystals, the orientation of each single-shot diffraction pattern was determined, from which a 3D diffraction pattern was generated using symmetry (Fig. 2d) (Methods). The 3D diffraction pattern was then iteratively phased to obtain a high-resolution image by incorporating symmetry into the oversampling smoothness (OSS) algorithm[32]. For each 3D diffraction pattern, we first performed 100 independent OSS reconstructions with random phases as their initial input, each consisting of 2,000 iterations and using a box as a loose support. For each independent reconstruction, symmetry was used as a constraint in real space and enforced once every 30 iterations (Methods). By averaging the five best reconstructions, we determined a tight support and performed another 100 independent OSS reconstructions, again enforcing symmetry once every 30 iterations. The five best reconstructions were averaged to obtain a final 3D image. Figures 3a-c, and Supplementary video S3a show iso-surface renderings of the final image along the 2-, 3- and 4-fold rotational symmetry, reconstructed from Fig. 2a. The concave facets of the nanocrystal are visible, and the insets show three 3.3-nm-thick centro-slices of the image. Using the same procedure, we obtained final 3D reconstructions from five other single-shot diffraction patterns (Supplementary Fig. S5). The overall shape, size and facets of the six reconstructions are in good agreement with each other (Fig. 2 and Supplementary Fig. S5). To investigate whether enforcing symmetry in the phase retrieval algorithm influenced the shape and structure of the reconstructed nanocrystals, we conducted numerical simulations on single-shot 3D structure determination of cubic, octahedral and trisoctahedral nanocrystals (all belonging to the point group $m\bar{3}m$ )[31]. Using simulated single-shot diffraction patterns (Supplementary Fig. S7) which have similar Poisson noise and missing data as the experimental patterns, we applied the same procedure for achieving single-shot 3D reconstructions. Supplementary Fig. S8 shows the 3D reconstructions of the cubic, octahedral and trisoctahedral nanocrystals, which are in good agreement with the 3D models (Supplementary Fig. S6). These results indicate that the shape, size and facets of the 3D reconstructions are intrinsically encoded in the single-shot diffraction patterns.

To evaluate the resolution of the six single-shot 3D reconstructions, we calculated the Fourier shell correction (FSC) between the reconstructions of the different nanocrystals (Methods). Figure 2c and Supplementary Fig. S9 show four representative FSC curves. Based on the FSC = 0.5 criterion, which has been widely used to estimate the resolution in single particle cryo-EM[27,33], we estimated the 3D resolution of these reconstructions to be ~5.5 nm. Next, we quantified the structure, shape and size of these reconstructed gold nanocrystals. Figure 4a and the inset show an iso-surface rendering and a 3.3-nm-thick central slice of a representative reconstructed nanocrystal along the 2-fold rotational symmetry, for which $D_1$, $D_2$, $D_3$, $\alpha$, $\beta$ and $\gamma$ are six parameters used to characterize the shape, size and facets of the nanocrystal. For each nanocrystal, we first determined the 3D orientations of the 24 facets and then measured the six parameters (Supplementary Table S1). The consistency of $D_1$, $D_2$ and $D_3$ among these six nanocrystals indicates that their overall shape and size are in good agreement. The three interfacial angles ($\alpha$, $\beta$ and $\gamma$) measured from all six nanocrystals agree with the values expected for a concave trisoctahedron with exposed {661} index facets (Supplementary Methods and Table S1). To further verify the {661} index facets, we calculated the interfacial angles of possible {$hhl$} index facets for a trisoctahedron (Supplementary Table S2), indicating that there are at least 1.4°, 1.7° and 1.9° differences for angles α, β and γ between {661} and other facets. As our six independent measurements of the three interfacial angles are more accurate than these angular differences (Supplementary Table S1), this further confirms that the concave trisoctahedral gold nanocrystals contain exposed high index {661} facets (Fig. 4b).

Next, we further examined whether the single-shot diffraction patterns were generated from identical gold nanocrystals by using cross-common arcs. Thirty pairs of cross-common arcs exist between six spherical diffraction patterns (Supplementary Methods). Supplementary Fig. S10 shows three representative pairs of the cross-common arcs between three single-shot diffraction patterns. The average $R_{arc}$ of the 30 pairs of cross-common arcs was calculated to be 9.2%, which is comparable to those obtained from the self-common arcs. After determining the relative orientations of the six single-shot patterns, we assembled a 3D diffraction pattern from 288 symmetrized 2D diffraction patterns (48 for each of the 6 nanocrystals). Using OSS with symmetry (Methods), we reconstructed the average 3D structure of the gold nanocrystals (Supplementary Fig. S11, and video S3b). The six parameters ($D_1$, $D_2$, $D_3$, α, β and γ) measured for the average structure are very consistent with a concave trisoctahedral nanocrystal containing exposed high index {661} facets (Supplementary Table S1). We also quantified the average 3D structure with respect to the single-shot 3D reconstructions by calculating FSC curves between them (Supplementary Fig. S12), indicating that the resolution of the average 3D reconstruction is presently limited by the detector size.

      While symmetry has long been used to determine the 3D structures of virus particles with cryo-EM[27,33], this work represents the first combination of symmetry and CDI for achieving single-shot 3D structure determination of nanocrystals at ~5.5 nm resolution using ~10 femtosecond X-FEL pulses. The significance of the present work is threefold. First, (to our knowledge) a single-shot 3D resolution of ~5.5 nm is the highest resolution ever achieved in any 3D X-ray imaging method. Second, as symmetry exists in many nanocrystals and virus particles[26,9,15,27], this method can be broadly applied to 3D structure studies of such particles at nanometer resolution on femtosecond time scales. Third, compared to cryo-EM[27,33], the intense, femtosecond X-FEL pulses and the high signal to noise ratio of the diffraction signal allow us to select single-shot diffraction patterns from highly symmetrical and identical nanocrystals based on self- and cross-common arcs (Methods). With a sufficient number of identical nanocrystals, this approach can in principle be used to determine the 3D structure of nanocrystals at atomic resolution. We anticipate that this work will find applications in coherent diffraction imaging, X-FEL science, materials science, nanoscience, chemistry, and structural biology.

## METHODS

**Data acquisition.** The single-shot X-FEL experiment was conducted using the multiple-application X-ray imaging chamber (MAXIC) installed in experimental hutch 3 (EH3) of SACLA. The X-ray energy was set at 5.4 keV with pulse energy of ~100 μJ. After bunch compression, the pulse duration of the X-FEL pulses was estimated to be ~10fs[29]. Gold nanocrystals, positioned randomly on a 100-nm-thick $Si_3N_4$ membrane, were exposed to single X-FEL pulses, which were focused to a ~1.5 μm spot by a pair of K-B mirrors[30]. The single-shot exposures were automated using a 2D scan program that controlled both the X-FEL pulse selector and the sample motor stage, ensuring only a single XFEL pulse impinged on the $Si_3N_4$ membrane at a programmed position. The sample was destroyed after exposure to a single X-FEL pulse, leaving a hole in the membrane (Supplementary Fig. S2). While the X-FEL pulses are delivered at a repetition rate of 10 Hz, the single-shot exposures can be performed at either 1 or 10 Hz. The present experiment was performed at 1 Hz. The single-shot X-ray diffraction patterns were recorded by an octal sensor, multiport CCD (octal-MPCCD) detector with 2048x2048 pixels and a pixel size of 50x50 $\mu m^2$. The octal-MPCCD was designed to allow an adjustable central aperture varying from complete closure to a maximum opening of 10x10 $mm^2$. The detector was placed at 1470 mm downstream of the sample, and the central aperture was set at 3x3 $mm^2$. The parasitic scatterings from the K-B mirrors as well as other X-ray optics were blocked by two sets of four-way cross slits inside the MAXIC; one set positioned close to the sample (at a distance of ~20 mm) and the other set further upstream of the sample (at a distance of ~350 mm). The experiment was performed under a vacuum pressure of ~$10^{-4}$ Pa within the MAXIC chamber.

**Data screening and post analysis.** Using the data acquisition scheme described above, we have collected terabytes of coherent diffraction patterns each generated by an X-FEL pulse. To detect 'true hits' within the collected data, we employed a data screening approach by comparing the average intensity in a pre-defined region of the detector with a threshold. By examining different thresholds heuristically, we were able to differentiate between no hits, partial hits, exact single particle hits, and multiple-particle hits. After extensive data screening, we obtained a total of 100 good quality diffraction patterns (Supplementary video S2). We then performed background subtraction and located the centre for each of the 100 diffraction patterns. To significantly enhance the signal-to-noise ratios of the diffraction intensity, we performed 13×13 binning for each pattern, which enhanced the signal to noise ratio and reduced the size of the diffraction patterns to 161×161 pixels.

Next, we applied a curvature correction to the diffraction patterns. Although the effect of the curvature of the Ewald sphere is small, it is not negligible in this experiment. The diffraction intensity extended to a highest resolution of ~4.6 nm, corresponding to a maximum diffraction angle ($2\theta_{max}$) of 2.87°, which was limited by the detector size. As each 2D diffraction pattern was collected by a planar CCD detector, we projected the diffraction intensity onto the Ewald sphere by re-sampling the intensity points[19]. After the curvature correction, we prioritized the 100 single-shot diffraction patterns by evaluating the self-common arcs and calculating an average $R_{arc}$ for each diffraction pattern (Supplementary Methods). The six single-shot diffraction patterns (Fig. 2a and Supplementary Fig. S3) with the smallest $R_{arc}$ were chosen for further analysis and reconstructions.

**Generating 3D diffraction patterns via symmetry.** Based on the self- and cross-common arcs for the six single-shot diffraction patterns (Supplementary Methods), we determined the orientation of each diffraction pattern with respect to a set of symmetry related primary axes (Fig. 2b). Using the symmetry operations, a 3D diffraction matrix on a Cartesian grid was generated from 48 symmetrized diffraction patterns. The value of a Cartesian point was calculated through linear interpolation of the neighbouring points available from the symmetrized diffraction patterns. Specifically, we assigned each Cartesian point at the origin of a sphere with a radius of 0.4 voxels, for which we defined the measured points within the sphere. We then interpolated the value of the Cartesian grid by

$$|F(\vec{k})| = \frac{\sum_{i=1}^{N} |F(\vec{k}_i)| \, w(\vec{k}, \vec{k}_i)}{\sum_{i=1}^{N} w(\vec{k}, \vec{k}_i)} \qquad and \qquad w(\vec{k}, \vec{k}_i) = \frac{1}{d(\vec{k}, \vec{k}_i)} \qquad (1)$$

where $|F(\vec{k})|$ and $|F(\vec{k}_i)|$ are the Fourier modulus of Cartesian grid $\vec{k}$ and measured point $\vec{k}_i$, $N$ is the total measured points within the sphere, and $d(\vec{k}, \vec{k}_i)$ is the distance between points $\vec{k}$ and $\vec{k}_i$. If $N = 0$, we assigned the value of the corresponding Cartesian point undefined. This 3D interpolation approach yielded seven 3D diffraction patterns, six of which were obtained from the individual single-shot diffraction patterns, and the seventh was assembled by combining the six patterns.

**Incorporation of symmetry into the oversampling smoothness (OSS) algorithm.** After verifying the existence of trisoctahedral symmetry in the single-shot diffraction patterns based on self-common arcs (Supplementary Methods), we incorporated this symmetry constraint into the OSS algorithm[32]. As the $m\bar{3}m$ point group is Fourier transform invariant[31], no interpolation is needed to enforce the symmetry operations during the phase retrieval process. To implement the symmetry constraint, we first located the voxels in the fundamental domain in real space (the shade area in Fig. 2b). We then applied the symmetry operations to these voxels to find the corresponding voxels in the other 47 domains. We averaged the symmetry related voxels in the 48 domains and assign each voxel the averaged value. For each 3D diffraction pattern, we performed 100 independent OSS reconstructions with a box as a loose support, where each reconstruction consisted of 2,000 iterations and the symmetry constraint was enforced once every 30 iterations. The oversampling ratio was 2.8 for the single-shot 3D diffraction patterns and 13.3 for the assembled 3D diffraction pattern. The figure of merit for each reconstruction was monitored by an $R_f$, defined as the difference between the experimental Fourier modulus and the calculated Fourier modulus in each iteration[32]. From each set of 100 reconstructions, we averaged the best five reconstructions with the smallest $R_f$, from which we determined a tight support. Using this improved support, we performed another set of 100 independent OSS reconstructions for each 3D diffraction pattern under the same procedure as the loose support case. The best five reconstructions were averaged to obtain the final 3D image for each diffraction pattern.

**Resolution estimation based on the Fourier shell correlation (FSC).** FSC has been widely used to estimate the 3D resolution in single particle imaging with cryo-EM[27,33]. Here we applied the FSC to estimate the resolution of six independent single-shot 3D reconstructions. Fifteen FSC curves were calculated between these 3D reconstructions. Supplementary Fig. S9 shows four representative FSC curves. Based on the FSC = 0.5 criterion, which has been used in single particle cryo-EM to estimate the resolution[33], we determined the resolution of the single-shot 3D reconstructions to be ~5.5 nm. We have also calculated six FSC curves between the 3D reconstruction of the assembled diffraction pattern and the single-shot 3D reconstructions. Supplementary Fig. S12 shows four representative FSC curves, indicating that the resolution of the 3D reconstruction of the assembled pattern is presently limited by the detector size.

## Acknowledgements

We thank H. Xia for providing monodisperse trisoctahedral gold nanocrystals used for this study, Z. H. Zhou and X. Zhang for stimulating discussions. This work was supported by the DARPA PULSE program through a grant from AMRDEC and the Institute of Physical and Chemical Research (RIKEN) of Japan. The experiments were performed at SACLA with the approval of JASRI and the program review committee (No. 2012A8027). H. J. acknowledges the support by the National Natural Science

Foundation of China (No. 51002089) and J.A.R. is supported by a Giannini Foundation postdoctoral fellowship.

**Figure legend**

**Figure 1**. Schematic layout of the single-shot 3D structure determination using femtosecond X-FEL pulses. X-FEL pulses were focused to a ~1.5 μm spot by a pair of K-B mirrors. Single-shot diffraction patterns were acquired from high-index-faceted gold nanocrystals based on the diffraction-before-destruction scheme. By using the symmetry of the nanocrystal, a 3D diffraction pattern was generated from a single-shot diffraction pattern, from which a 3D reconstruction was computed with a resolution of ~5.5 nm.

**Figure 2**. Symmetry verification of the trisoctahedral gold nanocrystals using the self-common arcs present in each single-shot diffraction pattern. **a**, A representative single-shot diffraction pattern. **b**, Stereographic projection of trisoctahedral symmetry, belonging to crystallographic point group $m\bar{3}m$ and consisting of 2- (ellipsis symbols), 3- (triangle symbols), 4-fold (diamond symbols) rotational symmetry and the centro-symmetry (white circles). **c**, A representative pair of self-common arcs, corresponding to two dark curves in (**a**). **d**, After verifying the trisoctahedral symmetry, a 3D diffraction pattern was assembled from 48 symmetrized copies of the single-shot diffraction pattern (**a**).

**Figure 3**. 3D reconstruction of the single-shot diffraction pattern (Fig. 2a). **a-c**, Iso-surface renderings of the final 3D reconstruction along the 2-, 3- and 4-fold rotational symmetry, respectively. The insets show 3.3-nm-thick centro-slices of the 3D reconstruction along the corresponding directions. The variation of the electron density inside the nanocrystal is within ±10%, which is mainly due to noise in the single-shot diffraction pattern. **d**, Fourier shell correction (FSC) comparison between two independently reconstructed gold nanocrystals. Based

on the FSC = 0.5 criterion, a 3D resolution of the reconstructions was estimated to be ~5.5 nm.

**Figure 4**. 3D model of a trisoctahedral gold nanocrystal constructed from a single-shot 3D reconstruction. **a**, Iso-surface rendering and 3.3-nm-thick central slice (inset) of the nanocrystal along the 2-fold rotational symmetry reconstructed from Fig. 2a, of which $D_1$, $D_2$, $D_3$, $\alpha$, $\beta$ and $\gamma$ are six parameters used to characterize the shape, size and facets of . **b**, 3D model of the concave trisoctahedral nanocrystal constructed from (**a**), containing exposed high index {661} facets (Supplementary Methods).

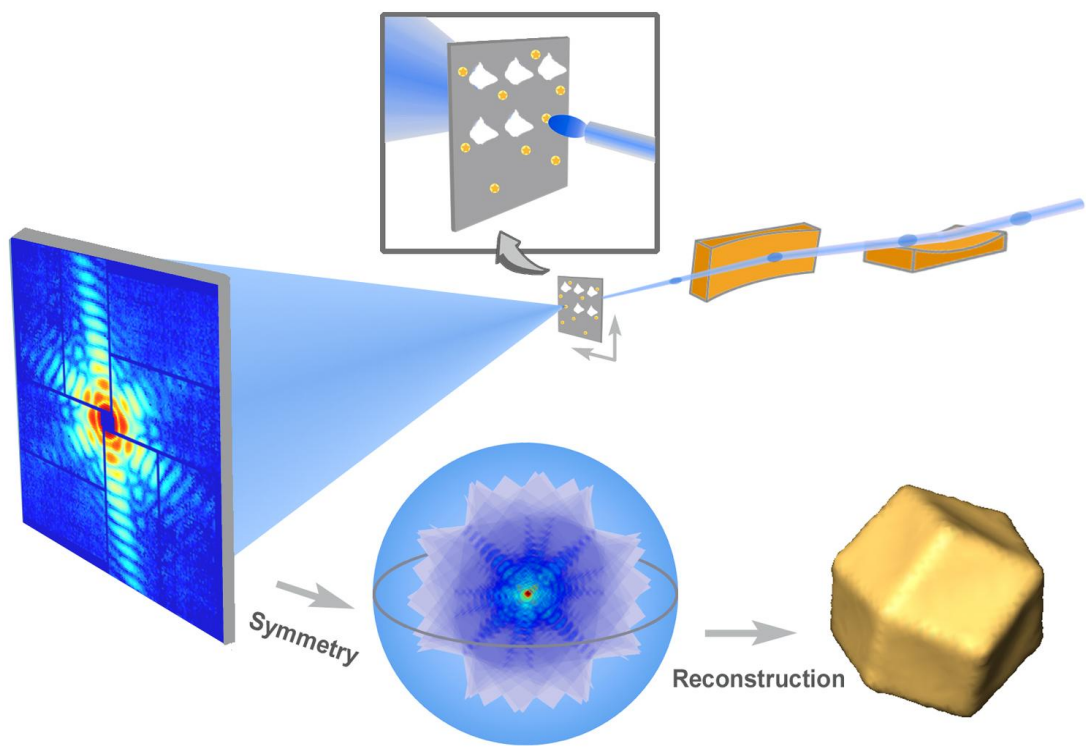

FIG. 1

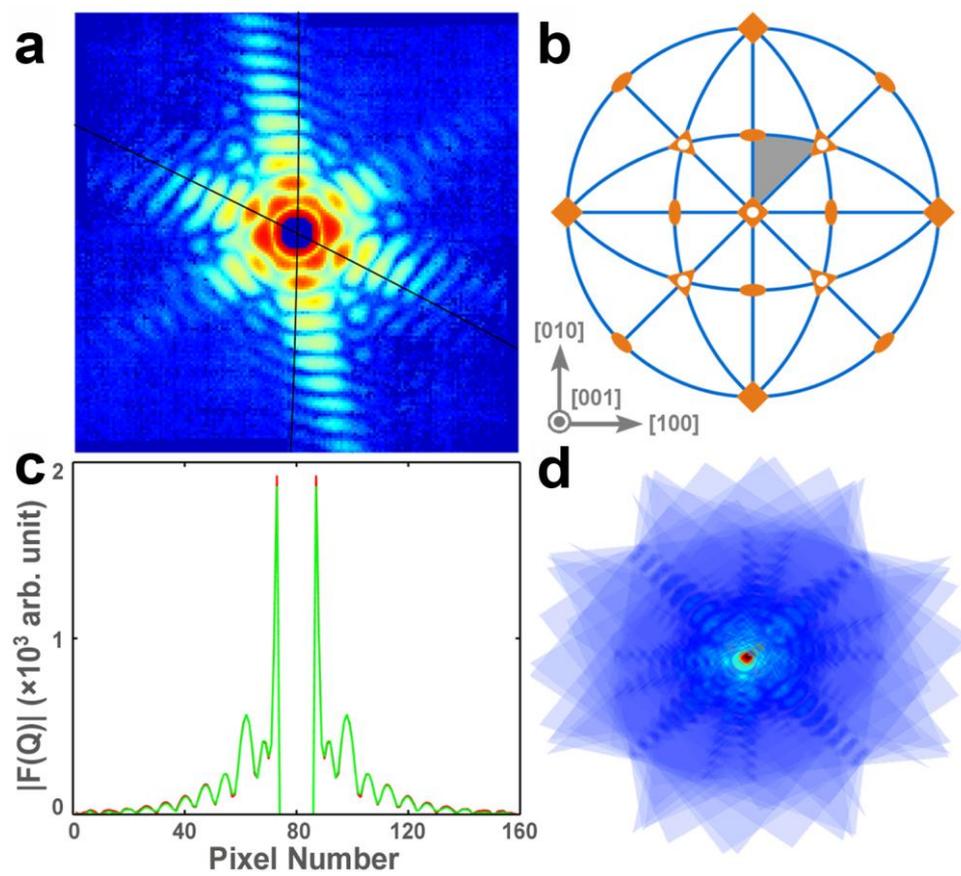

**a**

**b**

[010]

[001]

[100]

**c**

**d**

$|F(Q)|$ ($\times 10^3$ arb. unit)

0

0    40    80    120    160

**Pixel Number**

FIG. 2

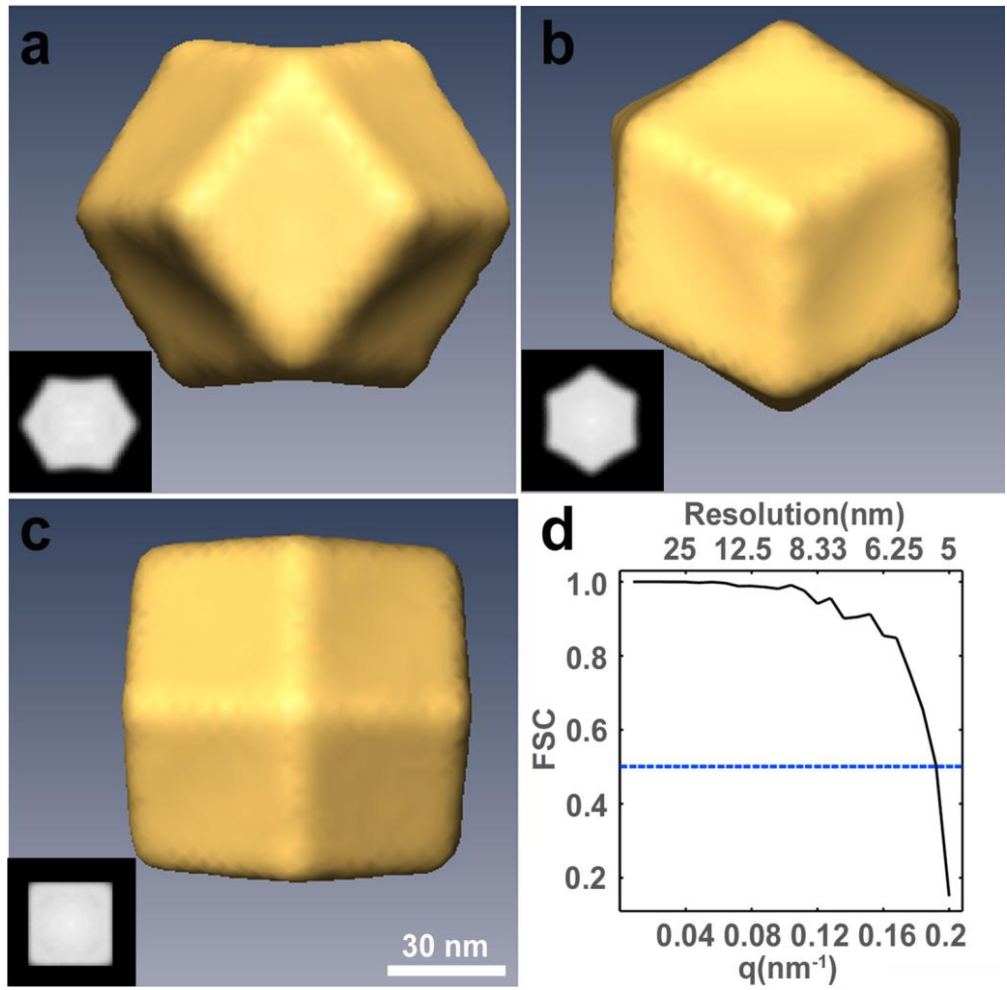

FIG. 3

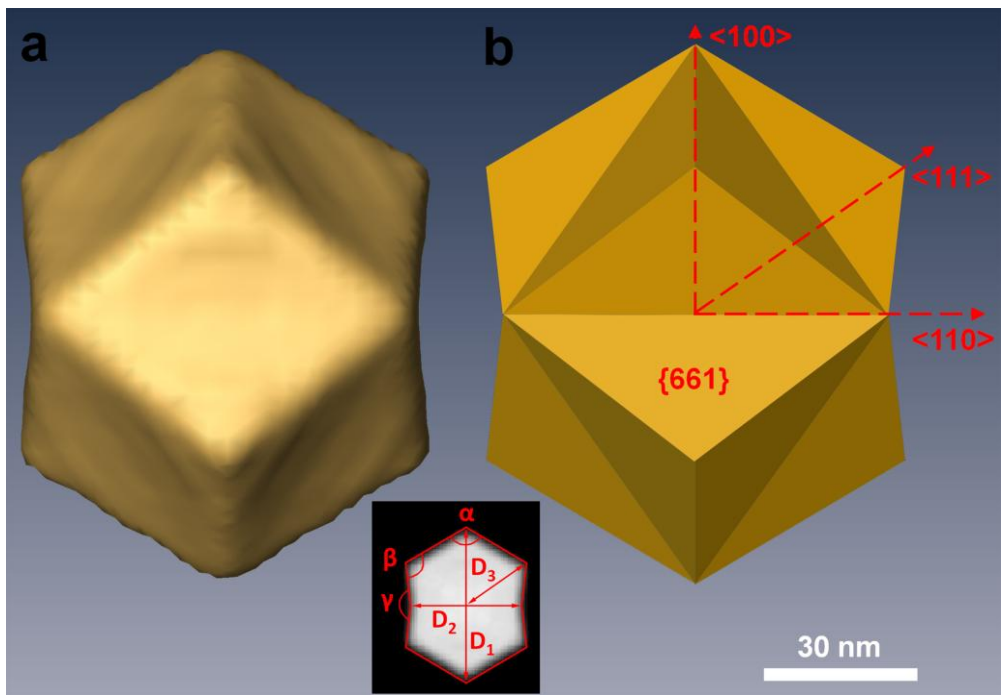

FIG. 4